\documentclass[%
 reprint,
 amsmath,amssymb,
 aps,
 prd,
]{revtex4-2}

\usepackage[svgnames]{xcolor}
\usepackage{graphicx}% Include figure files
\usepackage{dcolumn}% Align table columns on decimal point
\usepackage{bm}% bold math
\usepackage{hyperref}% add hypertext capabilities
\newcommand{\be}{\begin{equation}}

\newcommand{\ee}{\end{equation}}
\newcommand{\bea}{\begin{eqnarray}}
\newcommand{\eea}{\end{eqnarray}}

\def\f{\frac}
 
\newcommand{\lam}{\lambda}

\newcommand{\al}{{\alpha}}
\newcommand{\bet}{{\beta}}
\newcommand{\ga}{{\gamma}}
\newcommand{\om}{{ \omega}}

\newcommand{\bz}{{ \bar z}}
\newcommand{\bzeta}{{\bar\zeta}}

\newcommand{\idt}[1]{\int \widetilde{\rmd \vec #1}\, }
\def\rmd{{\rm d}}

\newcommand{\nn}{\nonumber}
\newcommand{\de}{\partial}

\usepackage[latin1]{inputenc}
\usepackage{color}

\begin{document}

\preprint{APS/123-QED}

\title{A celestial kinematical interpretation for an extended BMS$_4$}% Force line breaks with \\
%\thanks{A footnote to the article title}%
\author{Carles Batlle$^ {a)}$}\email{carles.batlle@upc.edu}
\author{Roberto Casalbuoni$^ {b)}$}\email{casalbuoni@fi.infn.it}
\author{Daniele Dominici$^ {b)}$}\email{dominici@fi.infn.it}
\author{José Figueroa-O'Farrill$^ {c)}$} \email{j.m.figueroa@ed.ac.uk}
\author{Joaquim Gomis$^ {d)}$} \email{joaquim.gomis@ub.edu}
\affiliation{$^ {a)}$ Institut d'Organitzaci\'o i Control i Departament de
	Matem\`atiques, Universitat Polit\`ecnica de Catalunya, EPSEVG,
	Av. V. Balaguer 1,  Vilanova i la Geltr\'u, 08800 Spain}
\affiliation{$^ {b)}$Department of Physics and Astronomy, University of Florence and
INFN, 50019 Sesto F.no (FI),  Italy}
\affiliation{$^ {c)}$Maxwell Institute and School of Mathematics, The University of Edinburgh, James Clerk Maxwell Building, Peter Guthrie Tait Road, King's Buildings, Edinburgh EH9 3FD, United Kingdom}
\affiliation{$^ {d)}$ Emeritus Professor of
%Departament de F\'isica Qu\`antica i Astrof\'isica 
%and Institut de Ci\`encies del Cosmos 
Universitat Barcelona,
Gran V\'ia de les Corts Catalanes 585,
08007 Barcelona,
Spain}

\date{\today}% It is always \today, today,
             %  but any date may be explicitly specified
\begin{abstract}
  Motivated by the work of Longhi and Materassi,
  who constructed a realisation of the (centreless) 
  BMS$_4$ algebra for the massive Klein-Gordon field in $3+1$, we
  build a realisation of the (centreless) massless BMS$_4$ algebra including
  super-rotations. %by using celestial  coordinates. 
  This realisation
  depends only on the momenta in the lightcone 
  %in the 
   expressed in celestial
  coordinates without any reference to the Klein--Gordon field. The
  quadratic Casimir of the Lorentz algebra  is written in terms of a
  second order differential operator and the volume form plays an
  essential role in this construction.  The BMS$_4$ algebra in terms of
  vector fields shows its kinematical nature, like the Poincar\'e
  algebra.  We also construct a dynamical realisation of BMS$_4$ from the
  symplectic structure on the solutions of the massless four-dimensional
  Klein--Gordon field in terms of quadratic expressions of the Fourier modes and plane waves invariant under translations. Using the Mellin transform,  we rewrite  the Klein--Gordon field in terms of the boost invariant basis, and write down the corresponding BMS$_4$ realization. We also provide the relation with spherical harmonics, linking our results with the solutions of Longhi-Materassi, which are in fact a subset of ours.
\end{abstract}
   %\pacs{02.20.-a; 02.20.Tw; 03.30.+p}

%\keywords{Suggested keywords}%Use showkeys class option if keyword
                              %display desired
\maketitle

%\tableofcontents

\section{Introduction}\label{sec:0}

The BMS$_4$ algebra was introduced in \cite{Bondi:1962px,Sachs:1962zza} as
the asymptotic symmetry algebra of a four-dimensional flat spacetime
at null infinity. An extended BMS$_4$ algebra which includes the Poincar\'e,
super-translations and super-rotations, was proposed by
\cite{Barnich:2009se}. A comprehensive presentation of recent
applications of BMS symmetries can be found in \cite{Strominger:2017zoo}.
Longhi and Materassi \cite{Longhi:1997zt} found a canonical
realisation of the BMS$_4$ algebra in terms of the natural symplectic
structure on the Fourier modes of a free massive Klein--Gordon (KG)
field in four-dimensional Minkowski spacetime,   extending
the translation algebra to include super-translations. In their
construction a fundamental role is played by one of the two Casimir
operators of the Lorentz algebra, through the general solution of
the eigenfunctions of the Casimir operator for a critical eigenvalue.

In the first part of this paper we extend this approach, by constructing a realization
of super-translations and super-rotations in the momentum space for the massless case. It is particularly convenient to make use of celestial coordinates, which play a crucial role in celestial holography (see the reviews \cite{raclariu2021lecturescelestialholography, 
Pasterski:2021rjz,Donnay:2023mrd,bagchi2025carrolliankaleidoscope} and references therein).
The  extended BMS algebra realisation, built in terms of differential operators of degrees zero and one, shows in this way its kinematic character,
 like the Poincar\'e algebra. This
construction should be valid for any free relativistic massless
theory. As an example we derive the BMS generators in the case of the
KG field, both in the plane-wave and boost basis. 

 We also study the relation of our solutions with  those obtained by Longhi and Materassi in terms of spherical harmonics, and it turns out that our construction yields more general solutions.

The rest of the paper is organized as follows. In Section \ref{sec:LorentzCC} we write the expression of the Lorentz generators in terms of light-cone celestial coordinates.
	Section \ref{sec:LCasimir} contains our main results, the construction of the generators of extended BMS$_4$ and the essential role of the quadratic Lorentz Casimir differential operator. Section \ref{sec:KG1} provides a realization of BMS$_4$ in terms of a massless Klein-Gordon field using the standard plane-wave representation, while the boost basis representation, related to the former by the Mellin transform, is presented in Section \ref{sec:KG2}. Section \ref{sec:LM} discusses the relation of our work with what would be the massless case of the Longhi-Materassi construction. We present our conclusions and outlook in Section \ref{sec:conc}. The appendixes present some technical results not included in the main text.

\section{Lorentz generators in celestial coordinates}
\label{sec:LorentzCC}

In this Section we will derive a kinematical representation of the
extended BMS algebra for a ligth-like momentum ($p^2=0$), by using the
celestial coordinates.  In this basis, the four momentum is given by  
\be
p^\mu=\om q^\mu,\,\,\,q^\mu=[1+z\bar z;z+\bz,-i(z-\bz), 1-z\bz]\,,
\label{momentabarz}
\ee
where $\omega$ is the lightcone energy
and $z,\, \bar z$ are the complex dynamical variables
that determine the momentum directions\footnote{The signature of the Minkowski metric is
  $(-+++)$}:
\be
\om=\f 1 2 (p^0+p^3),\quad \quad z=\f {p^1+ip^2}{p^0+p^3}.
\ee
On the other hand, in the Penrose description of Minkowski space-time, 
at the null boundary there is a Riemann sphere parameterized by complex
coordinates.  The starting point to construct celestial amplitudes is
the identification of the kinematical variables $z,\, \bar z$ as the
coordinates of this  sphere.  Note that the four-dimensional
Lorentz group, which is isomorphic to
$\operatorname{PSL}(2,\mathbb{C})$, acts on the celestial sphere via
conformal transformations (see, e.g., \cite{Stieberger:2018onx}).

In momentum space, the Lorentz generators 
\begin{equation}M_{0i}=-i p^0\de /{\de
  p^i},\quad  M_{ij}=i( p_i\de/{\de p^j}-p_j\de/ {\de p^i}),
 \end{equation}
    where
$p^0=\om (1+z\bz)$, can be re-expressed in celestial coordinates as %{\color{red}Note:  the sign of the $\partial_\omega$ term in $M_{02}$ was wrong.}
\bea
M_{01}&=&-\tfrac{i}2 [\om (z+\bz)\de_\om-(-1+z^2)\de_z -(-1+\bz ^2) \de_\bz],\nn\\
M_{02}&=&-\tfrac12 [\om (z-\bz)\de_\om-(1+z^2)\de_z +(1+\bz ^2) \de_\bz],\nn\\
M_{03}&=&i[z\de_z+\bz \de_\bz-\om \de_\om],
\label{eq:3}
\eea
and
\bea
M_{12}&=& -(z\de_z-\bz \de_\bz),\nn\\
M_{23}&=&-\tfrac12 [\om (-z+\bz)\de_\om +  (-1+z^2)\de_z- (-1+\bz^2)\de_\bz ],\nn\\
M_{31}&=&\tfrac{i}2 [-\om (z+\bz)\de_\om+(1+z^2)\de_z +(1+\bz ^2) \de_\bz].
\label{eq:4}
\eea

It is convenient to introduce the $L_n$ and $\bar L_n$ $n=0,\pm1$  operators defined as
\bea
L_0&=&\tfrac12 (M_{12}+i M_{03}),\nn\\
 L_{-1}&=& \tfrac12 (-M_{23}+iM_{31}-i M_{01}- M_{02}),\nn\\
\,L_{1}&=&\tfrac12 (M_{23}+iM_{31}+i M_{01}- M_{02}),\nn\\
\bar L_0&=&\tfrac12 (-M_{12}+i M_{03}),\nn\\
 \bar L_{-1}&=&\tfrac12 (M_{23}+iM_{31}-i M_{01}+ M_{02}),\nn\\
 \bar L_{1}&=&\tfrac12 (-M_{23}+iM_{31}+i M_{01}+ M_{02}).
\eea

With these definitions we get the explicit expressions
\bea
L_0&=&-z\de_z+\tfrac12 \om \de_\om,\nn\\
L_{-1}&=&-\de_z,
L_{1}=-z (z \de_z-\om \de_\om ),
\label{lm1}
\eea
and the corresponding ones for $\bar L_n$, $n=0,\pm 1$
which satisfy the following Lorentz algebra: 
\bea
&&[L_m,\,L_n ]=(m-n)\, L_{m+n},\,[\bar {L}_m,\,\bar {L}_n ]=(m-n)\,\bar {L}_{m+ n},\nn\\
&&[L_m,\,\bar {L}_n ]=0,\quad m,n=0,\pm 1.
\label{eq:9}
\eea

To consider the Poincar\'e algebra we introduce also the momenta,  as in \cite{Fotopoulos:2019vac}, 
\begin{align}
P_{-1/2,-1/2}=&P_0+P_3,\quad P_{-1/2,+1/2}=
P_1-i\,P_2,\nn\\
P_{1/2,-1/2}=&P_1+i\,P_2,\quad P_{1/2,1/2}=P_0-P_3.
\end{align}
In the representation given in eq.~(\ref{momentabarz}), we get
\begin{align}
P_{-1/2,-1/2}=& 2\omega,\quad P_{-1/2,+1/2}= 2\omega\bar z,\nn\\
P_{1/2,-1/2}=&2\omega z,\quad P_{1/2,1/2}=2\omega z\bar z.
\label{eq:20}
\end{align}

The Poincar\'e algebra  is satisfied
\begin{align}
[P_{i,j},\,P_{k,l}]=&0,\nn\\
[L_n,\,P_{k,l}]=& (\tfrac n 2 -k)P_{n+k,l},\nn\\
[\bar L_n,\,P_{k,l}]=& (\tfrac n 2 -l)P_{k,n+l},\quad i,j,k,l=\pm\tfrac12, n=0,\pm 1.
\label{bms}
\end{align}

\section{Lorentz Casimir and extended BMS algebra}
\label{sec:LCasimir}

\subsection{BMS algebra}

To build an extension of the Poincar\'e algebra containing
super-translations, we start, as shown in \cite{Longhi:1997zt}, from
the Casimir operator
\begin{equation}
  \label{casimir}
  \begin{aligned}
    C_2&=\sum_i (M_{0i}^2)-M_{12}^2-M_{23}^2-M_{31}^2\\
    &=-\om (3 \de_\om+\om \de^2_\om),
  \end{aligned}
\end{equation}
where we have used the explicit expression of the Lorentz generators
in celestial coordinates given in eqs.~(\ref{eq:3},\ref{eq:4}). The
second Casimir operator turns out to be zero, $\tfrac12
\epsilon_{\mu\nu\rho\sigma}M^{\mu\nu}M^{\rho\sigma}=0$.  Since we know
that a general four-vector is an eigenfunction of the $C_2$ operator
with eigenvalue -3, we consider all such eigenfunctions with the same
eigenvalue:
\begin{align}
C_2f(\om,z,\bz ) &=-\om (3 \de_\om+\om \de^2_\om) f(\om,z,\bz)\nn\\
&=-3 f(\om,z,\bz).
\label{casimireq}
\end{align}
Since the Casimir operator does not contain derivatives with 
respect to $z,\bz$, the general solution takes the form $f(\om,z,\bz)
= \omega g(z,\bz) + \omega^{-3} h(z,\bz)$, for some differentiable but
otherwise arbitrary functions $g(z,\bz)$ and $h(z,\bz)$.  Demanding
regularity sets $h(z,\bz) = 0$ and hence the regular solutions are given
by
\be
f(\om,z,\bz)= \om g(z,\bz)
\label{eq:35}
\ee
with $g(z,\bz)$ differentiable but otherwise arbitrary.

Since the solutions given in eq.~(\ref{eq:35}) are degenerate and since the commutator of $C_2$ with the helicity $M_{12}$ is zero, we now require that the solution be also eigenfunction of the helicity as in \cite{Longhi:1997zt}, or
\be
M_{12}g(z,\bz)=(z\de_z-\bz\de_\bz )g(z,\bz) = \lam g(z,\bz).
\ee
 Notice that the solution of the homogeneous equation 
\be\label{homogeneous}
M_{12}g(z,\bz)=0
\ee
is an arbitrary function of the modulus $|z|$,
since $M_{12}z=z, M_{12}\bar z=-\bar z$.  This means, since $M_{12}$ is a first-order differential operator, that we can multiply any solution corresponding to a given eigenvalue $\lambda$ by a function of $|z|$ without changing the eigenvalue. Taking this into account, the whole set of independent eigenfunctions of the Casimir with eigenvalue $-3$ and with definite eigenvalue of $M_{12}$ is spanned by functions of the form (\ref{eq:35}) with
 $g(z,\bz )= z^p\bz^q$, with $p,q\in \mathbb{Z}$, and which have $\lam=p-q$.

Let us define the super-translations as
\be
P_{-1/2+p,-1/2+q}(\om,z,\bz)=2\om z^p\bz^q,\quad p,q \in \mathbb{Z},
\label{momenta}
\ee
or equivalently
\be
P_{k,l}(\om,z,\bz) =2\om z^{k+1/2}\bz^{l+1/2},\quad k,l \in \mathbb{Z}+\f 1 2,
\label{momenta1}
\ee
with
\begin{eqnarray}
\lefteqn{M_{12}P_{-1/2+p,-1/2+q}(\om,z,\bz)}\nn\\&=&(p-q) P_{-1/2+p,-1/2+q}(\om,z,\bz).
\end{eqnarray}

The $P$'s in eqs.~(\ref{eq:20})  correspond to the choices $p,q=(0,0),(1,0),(0,1),(1,1)$. 
One can check that the BMS algebra given in eq.~(\ref{eq:9}) and
eq.~(\ref{bms}) is satisfied  for $k,l\in \mathbb{Z} + \tfrac12$.

\subsection{Extended BMS algebra}

Since the Lorentz generators are first-order differential operators we
consider the general holomorphic first-order differential operator
\be 
\xi=\xi^z (\om,z)\de_z+ \xi^\om(\om,z)\de_\om.
\ee
The condition $[C_2, \xi]=0$ implies
\be
\xi^z(\om,z)=\xi^z(z),\quad \xi^\om(\om,z)=\om C(z).
\label{eq:46b}
\ee
A second condition comes by requiring that the Lorentz-invariant
volume form
\begin{equation}
  \Omega=\om\ d\om\wedge  dz\wedge d\bz
\end{equation}
be invariant also under super-rotations, ${\cal L}_\xi \Omega= 0$, or
equivalently, due to
\be
{\cal L}_\xi \Omega= (\text{div}\ \xi) \,\, \Omega,
\ee
that $\xi$ be divergenceless.  Using Cartan's identity for the Lie
derivative ${\cal L}_\xi$  with respect to the vector field $\xi$ we
get
\be
(\de_\om +\f 1 \om)\xi^\om+\de_z \xi^z+\de_\bz \xi^{\bz}=0.
\label{eq:45}
\ee

Taking into account only the holomorphic part and plugging (\ref{eq:46b}) into (\ref{eq:45}) one gets the relation
\be
2 C(z)=-\de_z \xi^z(z).
\ee
If we consider the functions $\xi^z$ and $C$ to be expressed in Laurent series
\be
\xi^z(z)=\sum_{n}a_n z^n,\quad C(z)=\sum_{n}c_n z^n,
\ee
we obtain the general relation
\be
c_n=-\f {n+1} 2 a_n.
\ee
Using this, the generalization of the Lorentz generators $L_n, n=0,\pm 1$ for any $n\in \mathbb{Z}$ is given by
\be
L_n=-z^{n+1}\de_z+\f {n+1} 2 z^n\om\de_\om,\quad n\in \mathbb{Z},
\label{eq:21}
\ee
and similarly, for the anti-holomorphic part,
\be
\bar L_n=-\bz^{n+1}\de_\bz+\f {n+1} 2 \bz^n\om\de_\om,\quad n\in \mathbb{Z}.
\label{eq:22}
\ee

Note that the first term in the super-rotations is the standard
representation of the Witt algebra (centerless Virasoro algebra) in
two-dimensional conformal field theory.

The super-translation and super-rotations operators satisfy the
algebra given by eqs.~(\ref{bms}) for $m, n \in \mathbb{Z} $ and $i,
j, k, l \in  \mathbb{Z} + \tfrac12$.

We have then shown that the Poincar\'e algebra has an
infinite-dimensional extension  that we call extended BMS algebra.
This algebra has two interesting infinite-dimensional subalgebras: the
first one obtained extending the Poincar\'e algebra to BMS and the
second one extending the Lorentz algebra to the Witt algebra. The extended BMS algebra is the semi-direct sum of super-translations
and Witt.

Similar results were obtained by different methods by considering magnetic
Carrollian conformal scalar field theories in $2+1$ dimensions
\cite{Bekaert:2022oeh,Bekaert:2024itn,Chen:2024voz}.

\section{Klein--Gordon representation of extended BMS generators}
\label{sec:KG1}

Let us consider the expansion of a real Klein--Gordon field in the $z,\bz,\omega$ coordinates:
\be\label{field}
\phi(x)=\f 1 {(2\pi)^3}\int \om d\om d^2 z\left [ a(\om,z,\bz)e^{i\om q^\mu x_\mu }+h.c. \right ].
\ee
We can express the field also in Poincar\'e transformed variables
 \begin{align}
		x'^\mu  & = \Lambda^{-1}{}^\mu{}_\nu (x^\nu-b^\nu )\sim x^\mu -b^\mu -\lambda ^\mu {}_\nu x^\nu\,,\nonumber\\  p'_\mu =& (\Lambda^{-1})_\mu{}^\nu{} p_\nu=
		p_\nu \Lambda^\nu{}_\mu\sim p_\mu  +p_\nu\lambda^\nu {}_\mu
		\label{Poincparamatersconvention}
	\end{align}
	with $ p'_\mu x'^\mu= p_\mu( x^\mu-b^\mu )$.
Since $\phi$ is an scalar field
\bea
\phi(x)&=&\phi^\prime (x^\prime)\\
&=&\f 1 {(2\pi)^3}\int \om^\prime d\om^\prime d^2 z^\prime\left[ a^\prime(\om^\prime,z^\prime,\bz^\prime)e^{i\om^\prime q^{\mu^\prime} x_\mu^\prime }+h.c. \right],\nn
\eea
with the transformed variables  $\om', z', \bz'$ given by eqs.~(\ref{deltaom}),(\ref{deltaz}),(\ref{deltazb}).
Since the integration measure is invariant under these transformations 
\be
\om^\prime d\om^\prime d^2 z^\prime= \om d\om d^2 z,
\ee
we get
\bea
\phi(x) &=&\f 1 {(2\pi)^3}\int \om d\om d^2 z\left [ a(\om,z,\bz)e^{i\om q^\mu x_\mu }+h.c. \right ]\nn\\
%&=&\f 1 {(2\pi)^3}\int \om^\prime d\om^\prime d^2 z^\prime\left [ a^\prime(\om^\prime,z^\prime,\bz^\prime)e^{i\om^\prime q^{\prime\mu} x_\mu^\prime }+h.c. \right ]\nn\\
&=&\f 1 {(2\pi)^3}\int \om d\om d^2 z \Big [ a^\prime(\om^\prime,z^\prime,\bz^\prime)e^{i\om^\prime q^{\prime\mu} x_\mu^\prime }+h.c. \Big ]\nn\\
%&=&\f 1 {(2\pi)^3}\int \om d\om d^2 z \left [ a(\om^\prime,z^\prime,\bz^\prime)e^{i\om^\prime q^{\mu} x_\mu}e^{-i\om q^\mu b_\mu }+h.c. \right ]\nn\\
&\sim&\f 1 {(2\pi)^3}\int \om d\om d^2 z \Big [ a^\prime(\om^\prime,z^\prime,\bz^\prime)e^{i\om q^\mu x_\mu }(1-i\om  q^\mu b_\mu )\nn\\
\quad &+&h.c.\Big ],
\eea
where the $\sim$ in the last equation means that we have considered an infinitesimal transformation.
Making use of the result in \ref{appendixA}, we obtain
\be
\delta a (\om,z,\bz)= i\om  q^\mu b_\mu a (\om,z,\bz)-\f {\de a}{\de \om}\delta \om-\f {\de a}{\de z}\delta z-\f {\de a}{\de \bz}\delta \bz,
\label{deltaa}
\ee
which generalizes the result in (2.21) of \cite{Donnay:2022wvx} to include super-translations and super-rotations.

Using eqs.~(\ref{deltaom}),(\ref{deltaz}),(\ref{deltazb}), the infinitesimal variation of $\delta a$ can be rewritten in terms of the generators $P_{k,l}$, $L_n,\,\,n=-1,0,1$ as
\bea
\delta a(\om,z,\bz) &=& \big(\f i 2 b_{k,l}P_{k,l}+\beta L_{-1}+2 \alpha L_0  + \gamma L_1 \nn\\
&& +\bar\beta\bar L_{-1}+2\bar\alpha\bar L_0  +\bar\gamma\bar L_1 \big) a(\om,z,\bz),
\label{eq:87}
\eea
when $k,l=\pm{1/2}$ and where the translation parameters have been redefined as
\bea
b_{-1/2,-1/2}&=&b^3-b^0,\quad b_{1/2,1/2}=-b^0-b^3,\nn\\
b_{-1/2,1/2}&=&b^1+ib^2,\quad b_{1/2,-1/2}=b^1-ib^2 .
\eea

 The generalization of $\delta a(\om,z,\bz)$ to super-translations and super-rotations is straightforward:
\be
\delta a(\om,z,\bz) = \big(\f i 2 b_{k,l} P_{k,l}+ \epsilon_n L_{n}+\bar\epsilon_n\bar L_n \big)a(\om,z,\bz),
\ee
with $k,l \in  \mathbb{Z} + \f 12,n \in  \mathbb{Z}$, and
with a suitable choice of $\epsilon_n ,n=-1,0,1$ to reproduce eq.~(\ref{eq:87}).

In conclusion, assuming that the $\delta a(\om,z,\bz)$ is obtained, for example in the case of the super-translations, by making use of
\be
\delta a(\om,z,\bz)=b_{k,l}\{ {\cal P}_{k,l},a(\om,z,\bz)\},
\ee
 the generators of the extended BMS algebra  in the KG case are
\bea
  {\cal P}_{k,l} &= &\int \omega \,d\omega \,d^2 z \,P_{k,l} a^* (\omega, z,\bar z) a(\omega, z,\bar z))\,,\nonumber\\ 
 { \cal L}_{n}&=&  \int \omega \,d\omega \,d^2 z  a^* (\omega, z,\bar z) L_n  a(\omega, z,\bar z)\,,\nn\\ 
  { \bar {\cal L}}_{n}& =&\int \omega \,d\omega \,d^2 z a^*(\omega, z,\bar z){\bar L_n}a(\omega, z,\bar z),
 \label{bms0}
\eea
with $ k,l \in  \mathbb{Z} + \f 1 2,n \in  \mathbb{Z} $ and $P_{k,l},L_n,\bar L_n$ given by eqs.~(\ref{momenta1}),(\ref{eq:21}) and (\ref{eq:22}), if one assumes for the canonical Poisson bracket
\begin{align}
\lefteqn{\{a(\om,z,\bz),a^*(\om^\prime,z^\prime,\bz^\prime)\}}\nn\\ &=-i (2\pi)^3 \f 1 \om \delta(\om-\om^\prime)\delta^2(z-z^\prime) .
\end{align}

Note that the transformation of $a(\om,z,\bz)$ induces a transformation of the KG field via (\ref{field}) which is non-local except for the ordinary translations and Lorentz transformations.

An expression for the super-translations in terms of the Fourier modes of a Klein Gordon field was derived also in \cite{Banerjee:2018fgd}.

\section{Klein-Gordon field in the boost basis}
\label{sec:KG2}
The scalar field can be expanded in the boost basis as 
\begin{align}\label{DeltaExpansion} 
	\phi(x) &= \frac{1}{(2\pi)^{3}} \int d^2 z
	\int_{c-i\infty}^{c+i\infty} \frac{d\Delta}{i2\pi} \big[
	\phi^+_\Delta(x,z,\bz \,)a_{2-\Delta}(z,\bz\,)\nn\\
&\quad\quad\quad\quad \quad	\quad \quad +
	\phi^-_\Delta(x,z,\bz \,)a^\dagger_{2-\Delta}(z,\bz\,)
	\big],
\end{align}
where we have used 
the inverse Mellin transform formula 
\begin{equation}\label{InverseScalar}
	\int_{c-i\infty}^{c+i\infty}  \frac{\omega^{-\Delta}\,\Gamma(\Delta)}{(\epsilon \mp iq\cdot x)^\Delta} \frac{d\Delta}{i2\pi} = e^{\pm i\omega q\cdot x}\,,
\end{equation}
with $\epsilon >0$ a regulator to avoid the singularity at $q \cdot x=0$
and we have defined, 
as in  \cite{Pasterski:2021dqe},
\begin{align}
	a_\Delta(z,\bz \,) &= \int_0^\infty  \omega^{\Delta-1}a(\omega,z,\bz\,)\,d\omega,\label{adelta}\\ 
	a^\dagger_\Delta(z,\bz\,) &\equiv \int_0^\infty  \omega^{\Delta-1}a^\dagger(\omega\,z,\bz)\,d\omega\,.
	\label{adeltad}
\end{align}
They obey
\begin{align}
\lefteqn{\{a_\Delta(z,\bz \,),(a_{\Delta'}(z',\bz \,'))^\dagger\}}\nn\\
&=-i
(2\pi)^{4} \delta^2(z-z') 
\delta(i(\Delta+\Delta'^\ast-2)),
\end{align}
 with the generalized delta function defined  as in \cite{Donnay:2020guq}.

The wave functions $\phi^\pm_\Delta$ appearing in \eqref{DeltaExpansion} are the Mellin transforms of the plane waves 
\begin{equation}\label{conf0}
	\phi^\pm_\Delta(x;z,\bz)=
	\int_0^\infty d\omega \, \omega^{\Delta-1}\, e^{i(\pm q^\mu x_\mu+i\epsilon)\omega}\,.
\end{equation}
The integral in the rhs of \eqref{conf0} gives $\phi^\pm_\Delta=(\mp i)^\Delta\Gamma(\Delta)\varphi^\pm_\Delta$, where
\begin{equation}\label{varphi}
	\varphi^\pm_\Delta(x, z,\bz)
	=\frac{1}{(-q \cdot x_\pm )^\Delta},
\end{equation}
where $x^\mu_\pm=(x^0\mp i\epsilon, x^1,x^2,x^3)$. 
 Using the boost differential operators in space-time $M_{0i}=-x^0 \partial_i - x_i \partial^0$ and  that 
\begin{equation}
q^i M_{0i}(q^\mu x_\mu) =  -x^0 (q^i)^2 -x_i q^i q^0 = -q^0 q^\mu x_\mu,
\end{equation}
where $(q^i)^2=(q^0)^2$ has been used,
one can prove that
\begin{equation}
	q^i M_{0i} 	\phi^\pm_\Delta(x;z,\bz) = q^0 \Delta 	\phi^\pm_\Delta(x;z,\bz),
\end{equation}
and hence that the space-time functions $\phi^\pm_\Delta(x;z,\bz)$ are eigenfunctions of a boost transformation in the $q^i$ direction, with eigenvalue $q^0 \Delta$.

We can now compute, by using eq.~(\ref{deltaa}) and eqs.~(\ref{adelta})(\ref{adeltad}), the infinitesimal variation $\delta a_\Delta(z,\bz \,) $ 
\begin{align}
&\delta a_\Delta(z,\bz \,) = \int_0^\infty  d\om \omega^{\Delta-1}\delta a(\omega,z,\bz\,)
\nn\\
&=\int_0^\infty d\om  \omega^{\Delta-1}( \f i 2 P_{k,l}b_{k,l}+L_{n}\epsilon_n+\bar L_n \bar\epsilon_n)a(\om,z,\bz)\nn\\
&=i z^k\bz^l b_{k,l}   a_{\Delta+1}+\int_0^\infty d\om (L_{n}\epsilon_n+\bar L_n \bar\epsilon_n)a(\om,z,\bz).
\end{align}

Let us now evaluate
\begin{align}
&\int_0^\infty d\om L_{n} a(\om,z,\bz)\nn\\
&= \int_0^\infty d\om \om^{\Delta-1} \left (-z^{n+1}\de_z+\f {n+1} 2 z^n \om \de_\om \right )a(\om,z,\bz)\nn\\
&= -z^{n+1}\de_z a_{\Delta}+\f {n+1} 2 z^n \int_0^\infty d\om \om^{\Delta}   \de_\om a(\om,z,\bz)\nn\\
&= -z^{n+1}\de_z a_{\Delta}+\f {n+1} 2 z^n \int_0^\infty d\om \om^{\Delta} e^{-\epsilon\omega}\  \de_\om a(\om,z,\bz)\nn\\
&= -z^{n+1}\de_z a_{\Delta}- \Delta\f {n+1} 2 z^n \int_0^\infty d\om \om^{\Delta-1} e^{-\epsilon\omega} a(\om,z,\bz)\nn\\
&= -z^{n+1}\de_z a_{\Delta}- \Delta\f {n+1} 2 z^n a_\Delta (z,\bz),
\end{align}
where we have added a convergence term. Similar procedure can be followed  for the $\bar L_n$ term.
Summing up, the infinitesimal variation $\delta a_\Delta(z,\bz \,) $ under super-translations and super-rotations is
\bea
\delta a_\Delta(z,\bz \,) &=&i z^k\bz^l b_{k,l}   a_{\Delta+1}\nn\\
&-&\epsilon_n \left [z^{n+1}\de_z a_{\Delta}+ \Delta\f {n+1} 2 z^n a_\Delta (z,\bz)\right ]\nn\\
&-&\bar\epsilon_n \left [\bz^{n+1}\de_\bz a_{\Delta}+ \Delta\f {n+1} 2 \bz^n a_\Delta (z,\bz)\right ]
\eea
which, by identifying,
\be
{\cal T}=z^k\bz^l b_{k,l},\quad 
{\cal Y}^z=\epsilon_n z^{n+1},
\quad 
{\cal Y}^\bz=\bar\epsilon_n \bar z^{n+1}
\ee
and taking into account that for a scalar field $h=\bar h=\Delta$,
agrees with equation (4.50) of \cite{Donnay:2023mrd} and extends it to include super-translations and super-rotations.

\section{Relation with the construction in Longhi-Materassi}
\label{sec:LM}

In \cite{Longhi:1997zt}, Longhi and Materassi give an explicit expression of the canonical realization of  supertranslations for the massive case in terms of hypergeometric functions for the radial part and spherical harmonics for the angular dependence, and the relation of their expressions with the ones that we have obtained using the celestial coordinates should be investigated. Since we only have studied here the massless case, we will assume that the massless Longhi-Materassi construction has the same form than the massive one, replacing the complicated radial dependence by the simple form present in our solutions.

{
In order to make the connection more transparent, it is convenient to
use a different parameterization to the one in \eqref{momentabarz},
namely
\be
p^\mu=\varpi \pi^\mu,\,\,\,\pi^\mu=[1;\f{z+\bz}{1+z\bar z},\f{-i(z-\bz)}{1+z\bar z}, \f{1-z\bz}{1+z\bar z}]\,,
\ee
where
\be
\varpi =\om  (1+z\bar z).
\ee

In this case $(\pi^1)^2+(\pi^2)^2+(\pi^3)^2=1$ so that
$(\pi^1,\pi^2,\pi^3)$ are euclidean coordinates for points on the unit
sphere.  It will prove convenient to change coordinates.  To this end,
let us define
\be
\zeta=\f {z}{1+z\bar z}.
\ee
In these new variables,
\be
\pi^\mu=[1;{\zeta+\bzeta},{-i(\zeta-\bzeta)}, \sqrt{1-4\zeta\bzeta}]\
\ee
and we have
\bea
\de_{\om}& =& (1+z\bar z)\de_{\varpi},\nn\\
\de_{z}&= &\om \bz \de_{\varpi}+\f 1 {(1+z\bar z)^2}\de_{\zeta}-\f {\bz ^2} {(1+z\bar z)^2}\de_{\bzeta},\nn\\
\de_{\bz}&= &\om z \de_{\varpi}+\f 1 {(1+z\bar z)^2}\de_{\bzeta}-\f {z ^2} {(1+z\bar z)^2}\de_{\zeta}.
\eea
Changing coordinates from $(\omega,z,\bz)$ to $(\varpi,\zeta,\bzeta)$,
neither the second Casimir
\begin{equation}
  \label{casimir-2}
  \begin{aligned}
    C_2&=\sum_i M_{0i}^2-M_{12}^2-M_{23}^2-M_{31}^2\\
    &= -\om (3 \de_\om+\om \de^2_\om)\\
    &=-\varpi (3 \de_{\varpi}+{\varpi} \de^2_{\varpi}),
  \end{aligned}
\end{equation}
nor the rotation generator
\be\label{eq:J3}
M_{12} = z\de_z-\bz\de_\bz =\zeta\de_{\zeta}-{\bzeta}\de_{\bzeta} 
\ee
change form.
Therefore in the new variables the common eigenfunctions of $C_2$
(regular and with eigenvalue $-3$) and $M_{12}$ take the form
\be\label{supertrans}
\mathcal{P}_{p,q} := \varpi\zeta^p\bzeta^q f(|\zeta|)= \frac{\omega z^p \bz^q}{(1 + |z|^2)^{p+q-1}} f(\tfrac{|z|}{1+|z|^2})
\ee
with $p,q\in \mathbb{Z}$.
The dependence on the arbitrary differentiable function $f(|\zeta|)$
is due to the fact that it solves the homogeneous
equation~\eqref{homogeneous} and taking equation~\eqref{eq:J3} into
account.  These functions $\mathcal{P}_{p,q}$ can be related to the
$P_{k,l}$ in Section~\ref{sec:LorentzCC} by $P_{k,l} =
\mathcal{P}_{k+\frac12,l +\frac12}$.

We are now in a position to make contact with spherical harmonics.
Spherical harmonics are a basis for the space of square-integrable
functions on the sphere and this condition constraints the function
$f(|\zeta|)$ above.  Indeed, viewing the sphere as the unit sphere in 
three-dimensional euclidean space, spherical harmonics are the
restriction to the sphere of complex-valued homogeneous harmonic
polynomials in the euclidean coordinates.  More explicitly and up to
normalisation, the spherical harmonic $Y_{\ell m}$ is the restriction
to the sphere of a complex-valued homogeneous polynomial $\Pi_{\ell
  m}$ in $(\pi^1,\pi^2,\pi^3)$ of degree $\ell$, which is harmonic
\begin{equation}
  \label{eq:harmonic}
  \bigtriangleup \Pi_{\ell m} = 0,
\end{equation}
where $\bigtriangleup$ is the euclidean laplacian
\begin{equation}
  \label{eq:laplacian}
  \bigtriangleup = \partial_{\pi^1}^2 + \partial_{\pi^2}^2 + \partial_{\pi^3}^2,
\end{equation}
and such that $\Pi_{\ell m}$ is an eigenvalue of $M_{12}$
with eigenvalue $m \in \{-\ell,-\ell +1,\dots,\ell-1,\ell\}$:
\begin{equation}
  \label{eq:Plm-M}
  M_{12} \Pi_{\ell m} = m \Pi_{\ell m}.
\end{equation}
Homogeneous harmonic polynomials in $(\pi^1,\pi^2,\pi^3)$ of degree
$\ell$ are given by
\begin{equation}
  a_{i_1\dots i_\ell} \pi^{i_1} \dots \pi^{i_\ell},
\end{equation}
where $a_{i_1\dots i_\ell}$ is totally symmetric and
traceless\footnote{The dimension of the space of such
  polynomials is readily calculated to be
  $\dim\mathrm{Sym}^\ell(\mathbb{R}^3) - \dim
  \mathrm{Sym}^{\ell-2}(\mathbb{R}^3) = 2\ell + 1$, as expected.}:
\begin{equation}
  \delta^{ij} a_{ij i_1 \dots i_{\ell -2}} = 0.
\end{equation}
To impose the condition~\eqref{eq:Plm-M}, it is convenient to
re-express $\Pi_{\ell m}$ in terms of $\zeta = \pi^1 + i 
\pi^2$, $\bzeta = \pi^1 - i \pi^2$ and $\pi^3$, since these
coordinates diagonalise $M_{12}$:
\begin{equation}
  \label{eq:M12-diagonalised}
  M_{12} \zeta = \zeta, \quad M_{12}\bzeta = - \bzeta
  \quad\text{and}\quad M_{12} \pi^3 = 0.
\end{equation}

The euclidean laplacian in those coordinates is given by
\begin{equation}
  \label{eq:laplacian-zeta}
  \bigtriangleup = 4 \partial_{\zeta} \partial_{\bzeta} + \partial_{\pi^3}^2.
\end{equation}
The first few harmonic homogeneous polynomials are easy to write down:
\begin{equation}
  \begin{aligned}
    \Pi_{0 0} &= 1\\
    \Pi_{1 1} &= \zeta\\
    \Pi_{1 0} &= \pi^3\\
  \end{aligned}\qquad\qquad
  \begin{aligned}
    \Pi_{2 2} &= \zeta^2\\
    \Pi_{2 1} &= \zeta\pi^3\\
    \Pi_{2 0} &= \zeta\bzeta - 2 (\pi^3)^2,
  \end{aligned}
\end{equation}
with the convention that $\Pi_{\ell, -m} = \overline{\Pi}_{\ell m}$.
Their restriction to the sphere give the first few spherical
harmonics, up to normalisation.

In terms of the original coordinates $(z,\bz)$, the common
\emph{normalisable} eigenfunctions of $C_2$ (regular and with
eigenvalue $-3$) and $M_{12}$ take the form
\begin{equation}
  \mathcal{P}_{\ell m} = \varpi \Pi_{\ell m}(\zeta,\bzeta) = \omega (1 + |z|^2) \Pi_{\ell
    m}(\tfrac{z}{1+|z|^2},\tfrac{\bz}{1+|z|^2}),
\end{equation}
where we have already restricted the polynomials $\Pi_{\ell m}$ to the
sphere.  The first few are then
\begin{equation}
  \begin{aligned}
    \mathcal{P}_{0 0} &= \omega (1+|z|^2)\\
    \mathcal{P}_{1 1} &= \omega z\\
    \mathcal{P}_{1 0} &= \omega (1 - |z|^2)
  \end{aligned}\qquad
  \begin{aligned}
    \mathcal{P}_{2 2} &= \frac{\omega z^2}{1+|z|^2}\\
    \mathcal{P}_{2 1} &= \frac{\omega z (1-|z|^2)}{1 + |z|^2}\\
    \mathcal{P}_{2 0} &= \frac{\omega (|z|^2-2)(1-2|z|^2)}{2 (1 + |z|^2)}.
  \end{aligned}
\end{equation}
which, up to normalisation, are the massless analogues of the
solutions found by Longhi and Materassi in the massive case.

\section{Conclusions and outlook}
\label{sec:conc}

In this work we have seen that the extended BMS algebra can be realized in terms of differential operators of degrees zero and one in celestial coordinates on the lightcone, showing its kinematical nature, like the Poincar\'e group. As an example, we have built the realisation of 
the extended BMS algebra for a massless Klein--Gordon field. The same procedure should work for other massless free fields. The relation to the original construction of Longhi and Materassi, in terms of spherical harmonics, has also been investigated. Notice that the solutions 
(\ref{supertrans}) are more general than the ones corresponding to spherical harmonics, since the later are obtained with non-negative $p,q$. This is related to the fact that the spherical harmonics provide a basis for $L^2$ functions on the sphere, which is a requirement that has not been imposed in our construction.

One can try to  extend the construction described in this paper to the massive case, where the celestial coordinates are parameterized as
\be
p^\mu=\f m {2y}[1+y^2+z\bar z;z+\bz,-i(z-\bz), 1-y^2-z\bz].
\label{eq:4momenta}
\ee with inverse relation \be y=\f m {p^0+p^3},\quad z=\f {p ^1+ip^2}
{p^0+p^3}, \ee and Casimir operator given by \be C_2=-y^2\f \de {\de
  y^2} +y \f \de {\de y} - 4y^2 \de_z\de_\bz.  \ee The construction of
the super-translations in these coordinates should now recover the
original results of Longhi and Materassi, given in terms of
hypergeometric functions and spherical harmonics. However, if one
tries to construct super-rotations as first order differential
operators that commute with the Casimir and are divergenless as vector
fields, one has that the only solutions are the ordinary Lorentz
generators. This is due to the fact that any vector field commuting
with the Laplace-Beltrami operator associated to a metric is
necessarily Killing and the space of Killing vectors is bounded above
by $n(n+1)/2$ for an $n$-dimensional manifold.  The Lorentz Casimir
acting on functions is the Laplace-Beltrami operator on functions on
the mass hyperboloid (which is a hyperbolic space with $n=3$ in
$3+1$), and hence the only vector fields which commute with it are the
isometries of the mass hyperboloid (see, e.g.,
\cite[Proposition~4.2]{MR1395148}), which in this case is a
six-dimensional Lie algebra spanned by the Lorentz generators. Hence,
no super-rotations are allowed in the massive case. Problems with the
existence of super-rotations in the massive case in $2+1$ were also
pointed out in \cite{Batlle:2017llu}, using a different approach.

It would be interesting to understand the role of the Lorentz Casimir, which in our derivation is crucial, in the analysis of celestial amplitudes. Due to the isomorphism between extended BMS and conformal Carroll symmetries, a better understanding of the relation between our approach and  the magnetic Carrollian field theories in 2+1 dimensions \cite{Bekaert:2022oeh,Bekaert:2024itn,Chen:2024voz} deserves further study.

Another application of our approach would be to consider the $w_{1+\infty}$ algebra \cite{Strominger:2021mtt}, where the $\om$ independent part is already known \cite{Cappelli:1992yv}.

\hspace{1cm}
\begin{acknowledgments}
We acknowledge discussions with Andrea Campoleoni, Luca Ciambelli, Laura Donnay, Sabrina Pasterski and Antoine Van Proeyen.
One of the authors (JG) would like to acknowledge GGI for hospitality and partial financial support.
CB is partially supported by projects MAFALDA
(PID2021-126001OBC31, funded by MCIN/ AEI /10.13039/50110001 1033 and
by ``ERDF A way of making Europe''),  MASHED
(TED2021-129927B-I00, funded by MCIN/AEI /10.13039/501100011033 and by
the ``European Union Next GenerationEU/PRTR''), and ACaPE
(2021-SGR-00376, funded by AGAUR-Generalitat de Catalunya).
JG is partially funded by projects PID2022-136224NB-C21 (MICIU) and CEX2019-000918-M (ICCUB).
\end{acknowledgments}

\appendix

\section{Comment on the zero-modes of the on-shell Fourier transform}
\label{appendixA}

Let us first study the equation
\be
0=\idt p \left [ e^{i p^\mu x_\mu}f(\vec p)+ h.c\right ]
\label{A5}
\ee
by differentiating with respect to $x^0$ we get
\be
0=\idt p \left (-i p^0)[ e^{i p^\mu x_\mu}f(\vec p)- h.c\right ]
\ee
Let us multiply the first equation by $(ip^{\prime 0})e^{-i p^{\prime\mu} x_\mu}$ and the second by $e^{-i p^{\prime\mu} x_\mu}$, consider the difference and integrate over $d^3x$. We obtain
\begin{widetext}
\begin{align}
0&=\int d^3 x  \idt p \Big \{ ip^{\prime 0}\left [ e^{i (p^\mu -p^{\prime\mu})x_\mu}f(\vec p)+e^{-i (p^\mu +p^{\prime\mu})x_\mu}f^\dagger(\vec p) \right ]
+ ip^0  \left [ e^{i (p^\mu -p^{\prime\mu})x_\mu}f(\vec p)-e^{-i (p^\mu +p^{\prime\mu})x_\mu}f^\dagger(\vec p) \right ] \Big\}\nn\\
&=(2\pi)^3\idt p \Big \{ ip^{\prime 0}[\delta^3(p-p^\prime) e^{-i (p^0 -p^{\prime 0})x^0}f(\vec p) +\delta^3(p+p^\prime) e^{i (p^0 +p^{\prime 0})x^0}f^\dagger (\vec p)]\nn\\
&\quad\quad \quad\quad\quad\quad+ ip^{ 0}[\delta^3(p-p^\prime) e^{-i (p^0 -p^{\prime 0})x^0}f(\vec p) -\delta^3(p+p^\prime) e^{i (p^0 +p^{\prime 0})x^0}f^\dagger (\vec p)]\Big\}\nn\\
&= (2\pi)^3 i f({\vec p}^\prime)
\end{align}
\end{widetext}
where we have used
\be
\idt p=\int \f {d^3 p}{2 p^0}
\ee
and $p^0=\sqrt{\vec p^2}$.
In conclusion from eq. (\ref{A5}) we obtain
\be
f(\vec p)=0
\ee
This computation can be also repeated in the celestial basis, making use of 
\be
d^3p=2\om^2(1+\bz z)d\om d^2 z
\ee
and
\be
\delta^3(p^\prime -p)=\f 1 {2 \om^2(1+\bz z)}\delta({\om^\prime -\om}) \delta^2 (z^\prime -z).
\label{A1}
\ee
Then, from
\be
0=\int \om d\om d^2 z \left [ e^{i \om  q^\mu x_\mu}f(\om,z,\bz)+ h.c\right ],
\ee
one can deduce 
\be
f(\om,z,\bz)=0.
\ee

\section{Lorentz and $\operatorname{PSL}(2,\mathbb{C})$ transformations}
\label{app:B}
The generic matrix $M\in \operatorname{SL}(2,\mathbb{C})$ is written as
\be
M=\begin{pmatrix}
	a& -c\\
	-b& d\\
\end{pmatrix},
\ee
with the condition
$ \det M=ad-bc=1
$.
For an infinitesimal transformation we parametrize $M$ as
\be
M=\begin{pmatrix}
	1+\alpha& -\gamma\\
	-\beta& 1-\alpha\\
\end{pmatrix}.
\ee
$\operatorname{SL}(2,\mathbb{C})$ acts on the space of momenta (identified with the
space of hermitian $2\times 2$ matrices) via Lorentz transformations
$X \mapsto M X M^\dagger$.  Since $\pm I$ act trivially,  (the identity
component of) the Lorentz group is isomorphic to
$\operatorname{PSL}(2,\mathbb{C}) =
\operatorname{SL}(2,\mathbb{C})/\{\pm I\}$.

Using that isomorphism, we can compute the infinitesimal Lorentz transformations of $\om,z,\bz$:
\be
\delta\om=\f 1 2 (\delta p^0+\delta p^3)=\om [-(\al+\bar\al)+\ga z+\bar\ga \bar z]
\label{deltaom}
\ee
and
\be
\delta z= \delta \left (\f {p^1+i p^2}{p^0+p^3}\right )\sim2\al z-\ga z^2+\bet,
\label{deltaz}
\ee
that coincide with the results of \cite{Donnay:2020guq,Pasterski:2021rjz} when expanding the finite transformations for infinitesimal parameters.

In an analogous way we can obtain
\be
\delta \bz=2\bar\al \bz-\bar\ga \bz^2+\bar\bet.
\label{deltazb}
\ee

\bibliography{celestial}%

\end{document}